# The critical role of entropy in glass transition kinetics

Lijian Song[a,b], Meng Gao[a,b], Juntao Huo[a,b], Li-Min Wang[c]*, Yuanzheng Yue[d]*, Jun-Qiang Wang[a,b]*

[a] CAS Key Laboratory of Magnetic Materials and Devices, and Zhejiang Province Key Laboratory of Magnetic Materials and Application Technology, Ningbo Institute of Materials Technology and Engineering, Chinese Academy of Sciences, Ningbo 315201, China.

[b] Center of Materials Science and Optoelectronics Engineering, University of Chinese Academy of Sciences, Beijing 100049, China.

[c] State Key Lab of Metastable Materials Science and Technology, and College of Materials Science and Engineering, Yanshan University, Qinhuangdao, Hebei 066004, China.

[d] Department of Chemistry and Bioscience, Aalborg University, 9220 Aalborg, Denmark.

* Correspondence should be addressed to: jqwang@nimte.ac.cn; limin_wang@ysu.edu.cn; yy@bio.aau.dk

**Abstract**

Glass transition is a reversible transition that occurs in most amorphous materials. However, the nature of glass transition remains far from being clarified. A key to understand the glass transition is to clarify what determines the glass transition temperature ($T_g$) and liquid fragility ($m$). Here the glass transition thermodynamics for 150 different glass-forming systems are studied statistically. It is found that the activation characters in the energy landscape are crucial to precisely portray the glass transition and, in particular, both the activation free energy ($G^*$) and the activation entropy ($S^*$) play critical roles. $G^*$ determines $T_g$, $T_g=G^*/290+25.5$, while $S^*$ determines $m$, $m=S^*/R\ln 10+15$ with R is gas constant. Based on the Boltzmann definition of entropy, the fragility is an indication of the number of the degeneracy of the evolution paths. This explains why the nano-confined, low-dimension or high-pressured glasses exhibit stronger characteristics, which has been a puzzling phenomenon for a long time.

## 1. Introduction

Glass can form from various kinds of materials, including organic matters [1], inorganic compounds [2], alloys [3] and metal-organic frameworks [4], with only several exceptions like liquid helium and liquid nitrogen. The glass-liquid transition temperature ($T_g$) [5-10] is a measure of the thermal stability of glasses, i.e., the temperature limit of its elastic solid state, above which it transforms into a viscous liquid. Thus, it has been a long-lasting curiosity to design glasses with higher $T_g$ and to understand what determines $T_g$. It is reported that $T_g$ is closely related to the bond strength, elastic modulus and topological constraints [11-13]. A series of Ir- and Os-based bulk metallic glasses with very high $T_g$ have been designed in this strategy [3,14]. Ediger et al. [1,15] proposed to fabricate ultrastable glasses with enhanced $T_g$ through increasing the atomic bonding strength by surficial fast relaxation. Despite these experimental advancement in designing glasses with high $T_g$, the thermodynamic origin and kinetics of glass transition are far less understood.

The glass transition is a kinetic process that depends on the heating rate [7,16]. The rate dependence of $T_g$ is closely related with the fragility ($m$) of glass-forming liquids, which is a technical parameter defined by Angell [17] to quantify the non-Arrhenius behavior of a glass-forming liquid. The glasses with similar $T_g$ may have distinct $m$. For example, $T_g$ of the $Fe_{79}Si_{10}B_{11}$ metallic glass and $GeO_2$ glass are the same, i.e. 818 K, but their fragilities are significantly different, i.e. $m$=117 for $Fe_{79}Si_{10}B_{11}$ [18,19] versus $m$=20 for $GeO_2$ [20], respectively. The $m$ ranges from 15 to 214 for various glass-forming liquids [21,22]. The liquids with a small $m$ are usually called 'strong' liquids, whereas those with a large $m$ are called 'fragile' liquids. The physical properties of glasses have close relations with their $m$ values. For example, it is positively related to Poisson's ratio and plasticity [23]. A strong liquid usually exhibits higher glass forming ability [24,25] and a higher stability against crystallization [26,27]. Even though the technical definition for $m$ is clear and it has close correlations with other properties of glasses, the thermodynamic and structural origins of the liquid fragility remain less known.

In this work, based on the absolute reaction rate theory we calculate the thermodynamic parameters that characterize the glass transition of various glass systems, e.g., the activation enthalpy $H^*$, the activation free energy $G^*$ and the activation entropy $S^*$. We investigate the correlation between $T_g$ and $G^*$, as well as the relation between $S^*$ and the liquid fragility, so that the topographic complexity of potential energy landscape (PEL) can be revealed. Finally, we give experimental evaluations on the PEL of glass-forming liquids.

## 2. Methods

The two-state system is usually applied to describe the reaction kinetics. For example, the reaction rate is given as $k' \propto \exp\left(-\frac{H^*}{RT}\right)$, where $H^*$ is reaction barrier, $R$ is gas constant, $T$ is temperature. For glass transition, it is a process entailing multiple barriers. If there are $\Omega$ transition states/evolution paths with an averaged barrier of $H^*$, a factor of $\Omega$ should be multiplied with the single-barrier reaction rate, $k' \propto \exp\left(-\frac{H^*}{RT}\right) \times \Omega = \exp\left(-\frac{H^*}{RT}\right) \times \exp(\ln\Omega)$. Note that $\Omega$ is also regarded the thermodynamic probability of the state or the number of different ways in which a particular configuration can be achieved. Here, we define an activation entropy $S^*$ based on the Boltzmann definition of entropy,

$$S^* = R\ln\Omega \tag{1}$$

Then, it gives that

$$k' \propto \exp\left(-\frac{H^*}{RT}\right) \times \exp\left(\frac{S^*}{R}\right) \tag{2}$$

which is consistent with the *absolute reaction rate theory* (ART, also known as transition state theory, TST) [28-30], $-\frac{d\sigma}{dt} = \frac{\kappa k_B T}{h_P}\exp\left(-\frac{G^*}{RT}\right)\sigma = \frac{\kappa k_B T}{h_P}\exp\left(-\frac{H^*}{RT}\right)\exp\left(\frac{S^*}{R}\right)\sigma$, where $\sigma$ is the internal stress that is frozen-in during glass formation and is released during glass transition to supercooled liquid state, $t$ is time, $\kappa$ is the tunneling transmission coefficient with $\kappa = 1 + h_P\omega/(24k_BT)$, $\omega$ is angular frequency, $h_P$ is the Planck constant. $G^* = H^* - T_g S^*$ is activation free energy. $H^*$ represents a statistical averaged barrier for the transition states. It is linearly related to the kinetic energy barrier ($E^*$) determined in the Kissinger equation, $H^* = E^* - RT$ (see Supplementary Information or refs.[30,31]). During glass transition, the $d\sigma/dt$ reaches the maximum [32]. It gives $d(d\sigma/dt)/dt = 0$,

$$\ln\frac{R_h}{T^3} = \ln\frac{\kappa k_B R}{h_P} + \ln\frac{1}{H^*} + \frac{S^*}{R} - \frac{H^*}{RT} \tag{3}$$

$R_h$ is the heating rate. From Equation (3), the thermodynamic parameters ($H^*$, $S^*$) can be obtained by measuring the heating rate dependence of $T_g$.

## 3. Results

3.1 The kinetics characters of glass transition temperature

Figure 1 shows the relation between $H^*$ and $T_g$ for 150 different glass-forming systems, including metals, oxides, chalcogenides, molecules and polymers. The $T_g$ values of these systems range from 100 to 1000 K. In Fig. 1a, it is seen that the activation enthalpy $H^*$ increases along with $T_g$ in an approximately linear fashion, where the data are rather scattered. In Fig. 1b, there is no clear relation between $S^*$ and $T_g$. However, interestingly, when plotting the activation free energy $G^*$ ($=H^*-T_gS^*$) against $T_g$, we have observed a striking linear relation for those glass systems as shown in Fig. 1c. This suggests that $T_g$ is intimately associated with the activation free energy for glass transition. The slope of the linear relation is found to be about $G^*/T_g$=290 J mol$^{-1}$ K$^{-1}$. According to the elastic model [11,12,33,34], $G^*$ should be proportional to the atomic bonding strength.

Another intriguing point in Fig. 1c is that there is an intercept for the linear relation at about $T$=25.5 K at $G^*$=0. This implies that the lowest glass transition temperature should be above 25.5 K. This is supported by the experimental results. The lowest $T_g$ measured in experiments is 45.5 K for the glassy propane prepared by vapor deposition [35], while the second lowest $T_g$ is about 56 K obtained in glassy propene [36] and propylene [37]. The vapor-deposited amorphous ethane is expected to have a $T_g$ around 25.5 K, but it crystallizes at 23.9 K without exhibiting glass transition [35,37]. The correlation between boiling temperature and $T_g$ also suggests that $T_g$ may have a lower limit [38]. The missing of glass transition below 25.5 K is attributed to the quantum tunneling effect [39-41]. At low temperatures, the tunneling transmission coefficient in Eq. (1) $\kappa=1+\hbar_P\omega/24k_BT$ becomes large enough [39,40], which allows atoms to penetrate the small thermodynamic barrier of glasses with $T_g$ lower than 25.5 K. This explains why liquid nitrogen and liquid helium cannot be vitrified.

To examine the universality of the linear relation between $G^*$ and $T_g$, the $T_g$ values are compared between polyamorphisms in metallic glasses and between the as-cooled and the heavily-relaxed glasses as shown in Fig. 2. It is seen in Fig. 2a that the annealed glasses display higher $T_g$ than the as-quenched glasses. For NbNiZrTiCu high-entropy polyamorphic MG [42], $T_g$ increases from 672 K for the as-cooled sample to 732 K for the relaxed one, and correspondingly, $G^*$ increases from 190 to 210 kJ/mol (Fig. 2a). For CuZr-based MGs, $T_g$ increases from 651 to 685 K when $G^*$ increases from 186 to 190 kJ/mol (Fig. 2b). $T_g$ and $G^*$ values for Zr-based MGs were calculated based on the previously reported data [43]. It is seen that these values follow the same linear trend in Fig. 2b. This implies that the linear relationship between $G^*$ and $T_g$ is universal, i.e., independent of the type of chemical bonds and the polyamorphism of glasses.

3.2 Fragility and activation entropy

In Boltzmann definition of $S^*$ in Equation (1), a larger $S^*$ denotes more evolution paths or transition states (larger $\Omega$). The $\Omega$ reflects the degree of complexity in potential energy landscape (PEL). In the PEL theory, the complexity of energy landscape is thought to be related to the fragility index $m$ of the glass-forming liquid [7,44,45]. This motivates us to explore the relationship between $S^*$ and $m$. Strikingly and definitely, Fig. 3a shows that $m$ increases linearly with $S^*$ with a slop of about 0.050±0.03 $J^{-1}$ mol K, being consistent with the PEL theory [7,44]. The intercept ($S^*=0$) of the linear relation occurs at about $m=15$, i.e., at the lower limit of the liquid fragility, indicating that the activation entropy of the strongest glass-forming liquid (e.g., silica [20,46]) is close to zero.

The underlying physical mechanism for the linear relationship between $m$ and $S^*$ is studied based on the Mauro-Yue-Ellison-Gupta-Allan (MYEGA) model [21], $\log_{10}\eta(T) = \log_{10}\eta_\infty + (12 - \log_{10}\eta_\infty)\frac{T_g}{T}\exp\left[\left(\frac{m}{12-\log_{10}\eta_\infty} - 1\right)\left(\frac{T_g}{T} - 1\right)\right]$. The fragility index $m=\frac{d\log_{10}\eta(T)}{d(T_g/T)}$ and the activation energy $\frac{E^*}{R}$ can be linked by the temperature dependence of viscosity [47,48]: $E^* = mRT_g\ln10$ . Given $H^*=E^*-RT$ and the thermodynamic relation $S^* = \int \frac{\frac{dH^*}{dT}}{T} dT=(m - m_{\min})R\ln10$. It gives

$$m = S^*/R\ln10 + m_{\min} \tag{4}$$

where $m_{min}=15$ [49,50]. It is worth noting that the Volgel-Fulcher-Tammann (VFT) equation yields similar relationship (Eq. 4) with that from MYEGA model (see Supplementary data).

In Fig. 3a, the linear fitting with a slope of $1/R\ln10=0.052$ agrees well with the experimental data. The scattering of the experimental data should be attributed to the experimental errors of different measurements. Combining Eq. (1) and (4), we obtain the following equation, by which the number of the evolution paths can be estimated:

$$m = \log_{10}\Omega + 15 \quad (5)$$

Combining the relations $H^*=E^*-RT_g$, $E^*=mRT_g\ln10$ and $H^*=G^*+T_gS^*$, we obtain the relation $G^* = mRT_g\ln10 - RT_g - T_gS^*$, leading to the equation: $G^*/T_g = R(m_{min}\ln10-1)$. The slope $G^*/T_g$ in Fig. 1c determines the smallest fragility is about $m_{min}$=14.6±0.4, which is consistent with that reported previously [49]. To the best of our knowledge, this relation is the first to quantify the relation between dynamic liquid fragility and the thermodynamic probability of states (or the number of evolution paths of configurational states), as shown in Fig. 3b.

The number of the transition states $\Omega$ during glass transition should be related to the atomic/molecular motions, i.e. involve vibrational, translational, and rotating motions that can provide kinetic energy for transitions. This suggests that the activation entropy $S^*$ is positively related with vibrational entropy $S_{vib}$. When consider the degeneracy of transition states, $S^*$ should be inversely related with the configurational entropy $S_c$. The $S_{vib}$ can be obtained by measuring the vibrational density of states in experiments [51,52]. The ratio of $S_{vib}/S_c$ represents the degeneracy of transition states over static configuration. Figure 3c confirms the linear relation between $S^*$ and $S_{vib}/S_c$, which is also in line with refs. [53,54]. It should be mentioned that the present method is much easier to determine entropy compared to the neutron anelastic scattering experiments [51,52]. Either $S_c$ or $S^*$ can be obtained from calorimetry measurements. The $S_c$ can be calculated based on the heat capacity, while $S^*$ can be calculated based on the heating rate dependence of $T_g$.

To gain deeper insights into glass transition, the roles of $H^*$, $S^*$ and $G^*$ in glass forming process are illustrated in Fig. 4 based on the PEL. In PEL, $H^*$ represents the effective barrier for glass transition, which is a statistically averaged value of a big number of single barriers. It was generally accepted that a higher $H^*$ yields a higher $T_g$. Our results demonstrate that this effective barrier will decrease by $-TS^*$ if there are many evolution paths (number is $\Omega$). This yields a linear relation between $G^*$ ($=H^*-TS^*$) and $T_g$. As shown in Fig. 4a, $H^*$does not show linear relation with $S^*$, which breaks the Meyer-Neldel rule [55,56]. The contour lines of different $G^*=H^*-TS^*$ represent different $T_g$. As schematically illustrated in Fig. 4b, $S^*$ that is determined by the number of evolution paths determines the fragility of glass-forming liquid, while $G^*$ determines $T_g$.

## 4. Discussion

For glasses composed of atoms and simple molecules, the evolution paths are mainly composed of vibrational and translational motions. The degree of atomic motion freedom in 6 dimensions gives the $\Omega$ value of $(6.02\times10^{23})^6$, from which the liquid fragility is estimated to be 142.7. However, there is strong atomic bonding between each central atom and its nearest-neighbors or medium-range order range. The amount of the evolution paths or the liquid fragility should be smaller if stronger atomic bonds are involved in glass systems, as recently found by experiments [57].

For a glass-forming system with a rigid network, e.g., for the case of molten silica, its atom-evolution paths are depressed, and hence, its fragility approaches the lower limit. The molecular glass-forming liquids feature larger fragility than other systems for the following two reasons. First, the molecules exhibit rotational motions besides the vibrational and translational motions. Second, more low-frequency vibrations are generated in both liquids and solids, being driven by the increased soft/floppy modes. The degree of atomic motion freedom in 9 dimensions (3×vibrations, 3×translations and 3×rotations) gives the $\Omega$ value of $(6.02\times10^{23})^9$. These dynamic motions give an upper limit of the fragility, i.e., $m_{\max} = \log_{10}(6.02\times10^{23})^9+15=229$, which is slightly higher than the upper limit of $m \sim 175$ derived from the thermodynamic consideration [58] and the most fragile simulated silicon melt ($m \sim 200$) [59] and polyetherimide ($m$=214) [22].

It was found that liquid fragility is correlated with the heterogeneities of supercooled liquids and glasses [20,60]. It is also challenging to predict the stress relaxation behavior owing to the complex evolution of heterogeneity [61]. The mathematical definitions for activation entropy in Equation (1) and for fragility in Equation (5) based on the degeneracy of transition states may be helpful for establishing quantitative evaluation for heterogeneity in future.

The linear correlation between liquid fragility and activation entropy provides a crucial key to understand the connection between the dynamics and the thermodynamics in glass-forming liquids. Furthermore, it also allows us to explore the origin of the dependence of glass forming ability on liquid fragility. In detail, the higher the number of the evolution paths in a liquid system is, the higher is the probability for the system to lose its stability of the disordered structure, and the lower is the glass forming ability. This statement is consistent with previous observations [17,62]. Considering the liquid-glass inheritance [63], the glass derived from a fragile liquid is easier to deform, giving a higher Poisson's ratio [23] comparing with a strong liquid.

It has been long-lasting puzzling phenomena that the nano-confined, low-dimension or high-pressured glasses usually exhibit smaller fragility compared to bulk glasses under ambient pressure [64-66]. The correlation between fragility $m$ and $S^* = R\ln\Omega$ suggests that the nano-confinement, decrease in dimension or high-pressure can decrease the number of evolution paths. Thus, $S^*$ provides an importance route to design glasses with desirable properties.

The linear relation between $G^*$ and $T_g$ can be applied to estimate the free volume during glass transition. According to Hirai and Eyring [67], the fluidity of liquids is proportional to the probability for the atom to be adjacent to a hole and also to the rate at which it jumps into the hole. Based on this model [67], the shear viscosity of liquids can be expressed as $\eta_s = \frac{nN_0}{N_h}\frac{h_P}{v_0}\exp\left(\frac{H^* - TS^*}{RT}\right)$, where $v_0$ is the volume of atom ($\approx 2\times10^{-29}$ $m^3$) [11,68], $n$ is the ratio between volume of atoms and that of holes, $N_0$ and $N_h$ are the numbers of atoms and holes, respectively. Thus, the fraction ($N_h/nN_0$) of the hole volume (or the free volume) is estimated to be about ~4.7 % at $T_g$, which is consistent with literature [69].

## 5. Conclusions

In summary, we studied the activation free energy and activation entropy of the glass transition for various types of glasses based on the absolute reaction rate theory. The universal linear relations between thermodynamic and kinetic parameters for glass transition were discovered, i.e., the relations between the activation free energy $G^*$ and $T_g$, and between the activation entropy $S^*$ and the fragility index $m$ of glass forming liquids. The linear relationship between $G^*$ and $T_g$ suggests a lower $T_g$ limit of about 25.5 K, below which the quantum tunneling effect dominates and the statistical thermodynamic barrier for glass formation is negligible. According to the definition of the Boltzmann definition of entropy, we reveal that the liquid fragility is intimately correlated with the number of evolution paths, $m = \log_{10}\Omega + 15$. Our present work enables a quantitative analysis of glass transition in terms of the potential energy landscape and revealed the microscopic origin of liquid fragility.

## CRediT authorship contribution statement

**Lijian Song**: Writing–review & editing, Writing–original draft, Methodology, Funding acquisition, Formal analysis, Data curation. **Meng Gao**: Writing–original draft, Formal analysis. **Juntao Huo**: Writing–original draft, Formal analysis. **Li-Min Wang**: Writing–review & editing, Writing–original draft, Formal analysis. **Yuanzheng Yue**: Writing–review & editing, Writing–original draft, Formal analysis. **Jun-Qiang Wang**: Writing–review & editing, Writing– original draft, Supervision, Funding acquisition, Formal analysis, Conceptualization.

## Declaration of Competing Interest

The authors declare that they have no known competing financial interests or personal relationships that could have appeared to influence the work reported in this paper.

## Acknowledgments

This work was supported by the National Natural Science Foundation of China (52231006, 92163108, 52271158, 52222105, and 51827801), Zhejiang Provincial Natural Science Foundation of China (LGF22E010002, LZ22A030001, and LR22E010004) and "Pioneer and Leading Goose" R&D Program of Zhejiang (2022C01023)

## Supplementary materials

Supplementary material associated with this article can be found in the online version.

## References

[1] S.F. Swallen, K.L. Kearns, M.K. Mapes, Y.S. Kim, R.J. McMahon, M.D. Ediger, T. Wu, L. Yu, S. Satija, Organic glasses with exceptional thermodynamic and kinetic stability, Science 315 (2007) 353-356.

[2] G.N. Greaves, S. Sen, Inorganic glasses, glass-forming liquids and amorphizing solids, Adv. Phys. 56 (2007) 1-166.

[3] M.X. Li, S.F. Zhao, Z. Lu, A. Hirata, P. Wen, H.Y. Bai, M. Chen, J. Schroers, Y. Liu, W.H. Wang, High-temperature bulk metallic glasses developed by combinatorial methods, Nature 569 (2019) 99-103.

[4] R.S.K. Madsen, A. Qiao, J.N. Sen, I. Hung, K.Z. Chen, Z.H. Gan, S. Sen, Y.Z. Yue, Ultrahigh-field Zn-67 NMR reveals short-range disorder in zeolitic imidazolate framework glasses, Science 367 (2020) 1473-1476.

[5] P. Charbonneau, A. Ikeda, G. Parisi, F. Zamponi, Glass transition and random close packing above three dimensions, Phys. Rev. Lett. 107 (2011) 185702.

[6] H.B. Yu, R. Richert, R. Maass, K. Samwer, Unified criterion for temperature-induced and strain-driven glass transitions in metallic glass, Phys. Rev. Lett. 115 (2015) 135701.

[7] P.G. Debenedetti, F.H. Stillinger, Supercooled liquids and the glass transition, Nature 410 (2001) 259-267.

[8] S. Torquato, Glass transition: Hard knock for thermodynamics, Nature 405 (2000) 521-523.

[9] S. Sastry, The relationship between fragility, configurational entropy and the potential energy landscape of glass-forming liquids, Nature 40+ (2001) 164-167.

[10] Y.Z. Yue, C.A. Angell, Clarifying the glass-transition behaviour of water by comparison with hyperquenched inorganic glasses, Nature 427 (2004) 717-720.

[11] J.Q. Wang, W.H. Wang, Y.H. Liu, H.Y. Bai, Characterization of activation energy for flow in metallic glasses, Phys. Rev. B 83 (2011) 012201.

[12] W.H. Wang, The elastic properties, elastic models and elastic perspectives of metallic glasses, Prog. Mater. Sci. 57 (2012) 487-656.

[13] C. Hermansen, J.C. Mauro, Y. Yue, A model for phosphate glass topology considering the modifying ion sub-network, J. Chem. Phys. 140 (2014) 154501.

[14] J. Bi, X. Wei, X. Liu, R. Li, R. Xiao, T. Zhang, OsCo-based high-temperature bulk metallic glasses with robust mechanical properties, Scr. Mater. 228 (2023) 115336.

[15] K.J. Dawson, K.L. Kearns, L. Yu, W. Steffen, M.D. Ediger, Physical vapor deposition as a route to hidden amorphous states, Proc. Natl. Acad. Sci. 106 (2009) 15165-15170.

[16] M.S. Beasley, C. Bishop, B.J. Kasting, M.D. Ediger, Vapor-deposited ethylbenzene glasses approach "ideal glass" density, J. Phys. Chem. Lett. 10 (2019) 4069-4075.

[17] C.A. Angell, Formation of glasses from liquids and biopolymers, Science 267 (1995) 1924-1935.

[18] B.S. Dong, S.X. Zhou, J.Y. Qin, Y. Li, H. Chen, Y.G. Wang, The hidden disintegration of cluster heterogeneity in Fe-based glass-forming alloy melt, Prog. Nat. Sci. 28 (2018) 696-703.

[19] C. Zhang, Q. Chi, J. Zhang, Y. Dong, A. He, X. Zhang, P. Geng, J. Li, H. Xiao, J. Song, B. Shen, Correlation among the amorphous forming ability, viscosity, free-energy

difference and interfacial tension in Fe–Si–B–P soft magnetic alloys, J. Alloys Compd. 831 (2020) 154784.
[20] R. Böhmer, K.L. Ngai, C.A. Angell, D.J. Plazek, Nonexponential relaxations in strong and fragile glass formers, J. Chem. Phys. 99 (1993) 4201-4209.
[21] J.C. Mauro, Y.Z. Yue, A.J. Ellison, P.K. Gupta, D.C. Allan, Viscosity of glass-forming liquids, Proc. Nat. Acad. Sci. 106 (2009) 19780-19784.
[22] I. Echeverria, P.-C. Su, S.L. Simon, D.J. Plazek, Physical aging of a polyetherimide: creep and DSC measurements, J. Polym. Sci. B Polym. Phys. 33 (1995) 2457-2468.
[23] V.N. Novikov, A.P. Sokolov, Poisson's ratio and the fragility of glass-forming liquids, Nature 431 (2004) 961–963.
[24] W.L. Johnson, J.H. Na, M.D. Demetriou, Quantifying the origin of metallic glass formation, Nat. Commun. 7 (2016) 10313.
[25] R. Busch, E. Bakke, W.L. Johnson, Viscosity of the supercooled liquid and relaxation at the glass transition of the $Zr_{46.75}Ti_{8.25}Cu_{7.5}Ni_{10}Be_{27.5}$ bulk metallic glass forming alloy, Acta Mater. 46 (1998) 4725-4732.
[26] Y. Zhao, B.S. Shang, B. Zhang, X. Tong, H.B. Ke, H.Y. Bai, W.-H. Wang, Ultrastable metallic glass by room temperature aging, Sci. Adv. 8 (2022) eabn3623.
[27] J. Orava, A.L. Greer, Fast and slow crystal growth kinetics in glass-forming melts, J. Chem. Phys. 140 (2014) 214504.
[28] F.W. Cagle, H. Eyring, An application of the absolute reaction rate theory to some problems in annealing, J. Appl. Phys. 22 (1951) 771-775.
[29] H. Eyring, The theory of absolute reaction rates, Trans. Faraday Soc. 34 (1938) 41-48.
[30] L.J. Song, W. Xu, J.T. Huo, F.S. Li, L.-M. Wang, M.D. Ediger, J.-Q. Wang, Activation entropy as a key factor controlling the memory effect in glasses, Phys. Rev. Lett. 125 (2020) 135501.
[31] H.W. Starkweather, Simple and complex relaxations, Macromolecules 182 (1981) 1277-1281.
[32] Q. Yang, S.-X. Peng, Z. Wang, H.-B. Yu, Shadow glass transition as a thermodynamic signature of β relaxation in hyper-quenched metallic glasses, Nat. Sci. Rev. 7 (2020) 1896-1905.
[33] J.C. Dyre, Colloquium: The glass transition and elastic models of glass-forming liquids, Rev. Mod. Phys. 78 (2006) 953-972.
[34] J.C. Dyre, W.H. Wang, The instantaneous shear modulus in the shoving model, J. Chem. Phys. 136 (2012) 224108.
[35] K. Takeda, M. Oguni, H. Suga, A DTA apparatus for vapour-deposited samples. characterisation of some vapour-deposited hydrocarbons, Thermochim. Acta 158 (1990) 195-203.
[36] O. Haida, H. Suga, S. Seki, Realization of the glassy state of some simple liquds by the vapor condensation method, Thermochim. Acta 3 (1972) 177-180.
[37] I. Sugawara, Y. Tabata, The relationship between phase transition and luminescence in simple hydrocarbons, Chem. Phys. Lett. 41 (1976) 357-361.
[38] L.-M. Wang, R. Richert, Glass transition dynamics and boiling temperatures of molecular liquids and their isomers, J. Phys. Chem. B 111 (2007) 3201–3207.

[39] R.T. Skodje, D.G. Truhlar, Parabolic tunneling calculations, J. Phys. Chem. 85 (1981) 624-628.
[40] I.R. Sims, Low-temperature reactions: Tunnelling in space, Nat. Chem. 5 (2013) 734-736.
[41] V.N. Novikov, A.P. Sokolov, Role of quantum effects in the glass transition, Phys. Rev. Lett. 110 (2013) 065701.
[42] H. Luan, X. Zhang, H. Ding, F. Zhang, J.H. Luan, Z.B. Jiao, Y.C. Yang, H. Bu, R. Wang, J. Gu, C. Shao, Q. Yu, Y. Shao, Q. Zeng, N. Chen, C.T. Liu, K.F. Yao, High-entropy induced a glass-to-glass transition in a metallic glass, Nat. Commun. 13 (2022) 2183.
[43] Q. Sun, D.M. Miskovic, K. Laws, H. Kong, X. Geng, M. Ferry, Transition towards ultrastable metallic glasses in Zr-based thin films, Appl. Surf. Sci. 533 (2020) 147453.
[44] F.H. Stillinger, A topographic view of supercooled liquids and glass formation, Science 267 (1995) 1935.
[45] A. Thirumalaiswamy, R.A. Riggleman, J.C. Crocker, Exploring canyons in glassy energy landscapes using metadynamics, Proc. Natl. Acad. Sci. 119 (2022) e2210535119.
[46] Y.Z. Yue, Anomalous enthalpy relaxation in vitreous silica, Front. Mater. 2 (2015) 1-11.
[47] A.Q. Tool, Relaxation between inelastic deformability and thermal expansion of glass in its annealing range, J. Am. Ceram. Soc 29 (1946) 240-253.
[48] D.J. Plazek, K.L. Ngai, Correlation of polymer segmental chain dynamics with temperature-dependent time-scale shifts, Macromolecules 24 (1991) 1222-1224.
[49] Q. Zheng, J.C. Mauro, A.J. Ellison, M. Potuzak, Y. Yue, Universality of the high-temperature viscosity limit of silicate liquids, Phys. Rev. B 83 (2011) 212202.
[50] H.S. Chen, A method for evaluating viscosities of metallic glasses from the rates of thermal transformations, J. Non-Cryst. Solids 27 (1978) 257-263.
[51] A. Zaccone, Relaxation and vibrational properties in metal alloys and other disordered systems, J. Phys. Condens. Matter. 32 (2020) 203001.
[52] H.L. Smith, C.W. Li, A. Hoff, G.R. Garrett, D.S. Kim, F.C. Yang, M.S. Lucas, T. Swan-Wood, J.Y.Y. Lin, M.B. Stone, D.L. Abernathy, M.D. Demetriou, B. Fultz, Separating the configurational and vibrational entropy contributions in metallic glasses, Nat. Phys. 13 (2017) 900-905.
[53] L.-M. Wang, R. Richert, Measuring the configurational heat capacity of liquids, Phys. Rev. Lett. 99 (2007) 185701.
[54] P.F. Li, P. Gao, Y.D. Liu, L.-M. Wang, Melting entropy and its connection to kinetic fragility in glass forming materials, J. Alloys Compd. 696 (2017) 754-759.
[55] Y.-J. Wang, M. Zhang, L. Liu, S. Ogata, L.H. Dai, Universal enthalpy-entropy compensation rule for the deformation of metallic glasses, Phys. Rev. B 92 (2015).
[56] W. Meyer, H. Neldel, Relation between the energy constant ε and the quantity constant α in the conductivity-temperature formula of oxide semiconductors, Z. Tech. Phys. 18 (1937) 588-593.
[57] Y. Shi, B. Deng, O. Gulbiten, M. Bauchy, Q. Zhou, J. Neuefeind, S.R. Elliott, N.J. Smith, D.C. Allan, Revealing the relationship between liquid fragility and medium-range order in silicate glasses, Nat. Commun. 14 (2023) 13.

[58] L.-M. Wang, C.A. Angell, R. Richert, Fragility and thermodynamics in nonpolymeric glass-forming liquids, J. Chem. Phys. 125 (2006) 074505.
[59] V. Molinero, S. Sastry, C.A. Angell, Tuning of tetrahedrality in a silicon potential yields a series of monatomic (metal-like) glass formers of very high fragility, Phys. Rev. Lett. 97 (2006) 075701.
[60] A.S. Manz, M. Aly, L.J. Kaufman, Correlating fragility and heterogeneous dynamics in polystyrene through single molecule studies, J. Chem. Phys. 151 (2019) 084501.
[61] H.N. Lee, K. Paeng, S.F. Swallen, M.D. Ediger, Direct measurement of molecular mobility in actively deformed polymer glasses, Science 323 (2009) 231-234.
[62] H. Tanaka, Relationship among glass-forming ability, fragility, and short-range bond ordering of liquids, J. Non-Cryst. Solids 351 (2005) 678-690.
[63] A. Widmer-Cooper, P. Harrowell, H. Fynewever, How reproducible are dynamic heterogeneities in a supercooled liquid? , Phys. Rev. Lett. 93 (2004) 135701.
[64] C. Zhang, Y. Guo, K.B. Shepard, R.D. Priestley, Fragility of an isochorically confined polymer glass, J. Phys. Chem. Lett. 4 (2013) 431-436.
[65] L. Cao, L.J. Song, Y.R. Cao, W. Xu, J.T. Huo, Y.Z. Lu, J.-Q. Wang, Small activation entropy bestows high-stability of nanoconfined D-mannitol, Chin. Phy. B 30 (2021) 076103.
[66] C.M. Roland, S. Hensel-Bielowka, M. Paluch, R. Casalini, Supercooled dynamics of glass-forming liquids and polymers under hydrostatic pressure, Rep. Prog. Phys 68 (2005) 1405-1478.
[67] N. Hirai, H. Eyring, Bulk viscosity of polymeric systems, J. Polym. Sci. 37 (1959) 51-70.
[68] C.H. Reichardt, G. Halsey, H. Eyring, Mechanical properties of textiles, 10: analysis of steinberger's data on creep of cellulose acetate filaments, Text. Res. J. 16 (1946) 382-389.
[69] H.B. Ke, P. Wen, D.Q. Zhao, W.H. Wang, Correlation between dynamic flow and thermodynamic glass transition in metallic glasses, Appl. Phys. Lett. 96 (2010) 251902.
[70] Y.R. Cao, F.R. Wang, L.J. Song, M.Z. Li, A. Li, J.T. Huo, H. Li, F.S. Li, P. Yu, W. Xu, J.-Q. Wang, The nanocopper interface induces the formation of a new ultrastable glass phase, J. Non-Cryst. Solids 593 (2022) 121764.

## Figures and Tables

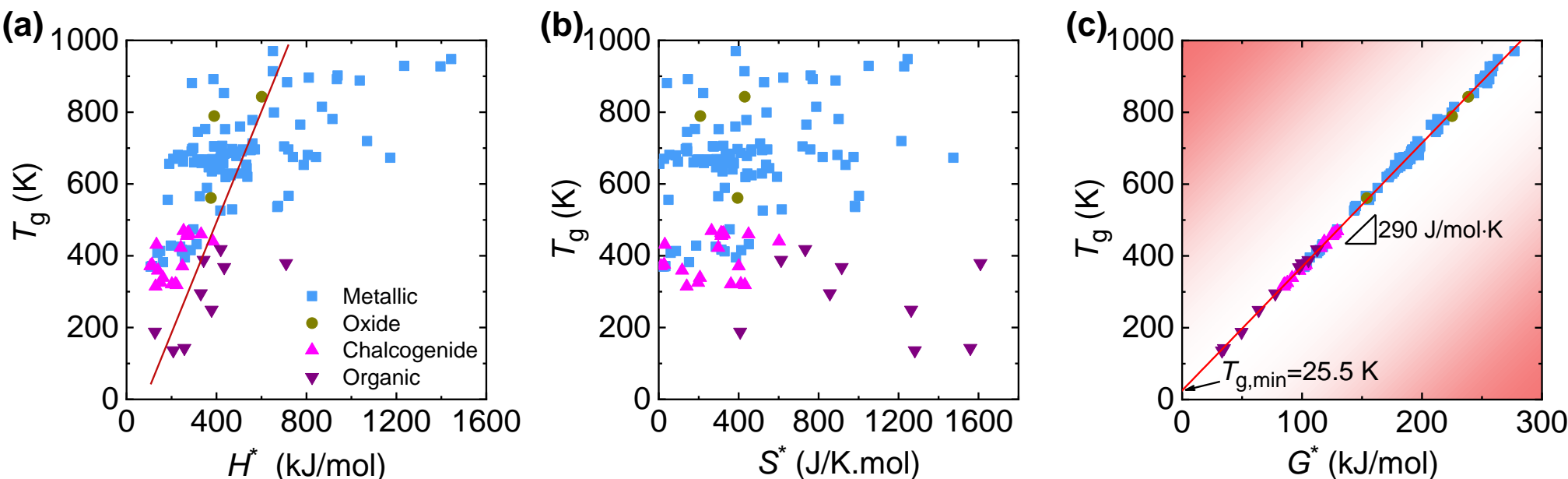

Fig. 1. The thermodynamic parameters versus glass transition temperature $T_g$ for different types of glasses. (a) The activation enthalpy $H^*$ versus $T_g$ for metallic glasses, organic glasses, oxide glasses and chalcogenide glasses. (b) The activation entropy $S^*$ versus $T_g$. (c) The activation free energy $G^*$ versus $T_g$. The linear fitting yields a slope of 290 J $mol^{-1}$ $K^{-1}$ and an intercept of 25.5 K (lower limit of $T_g$).

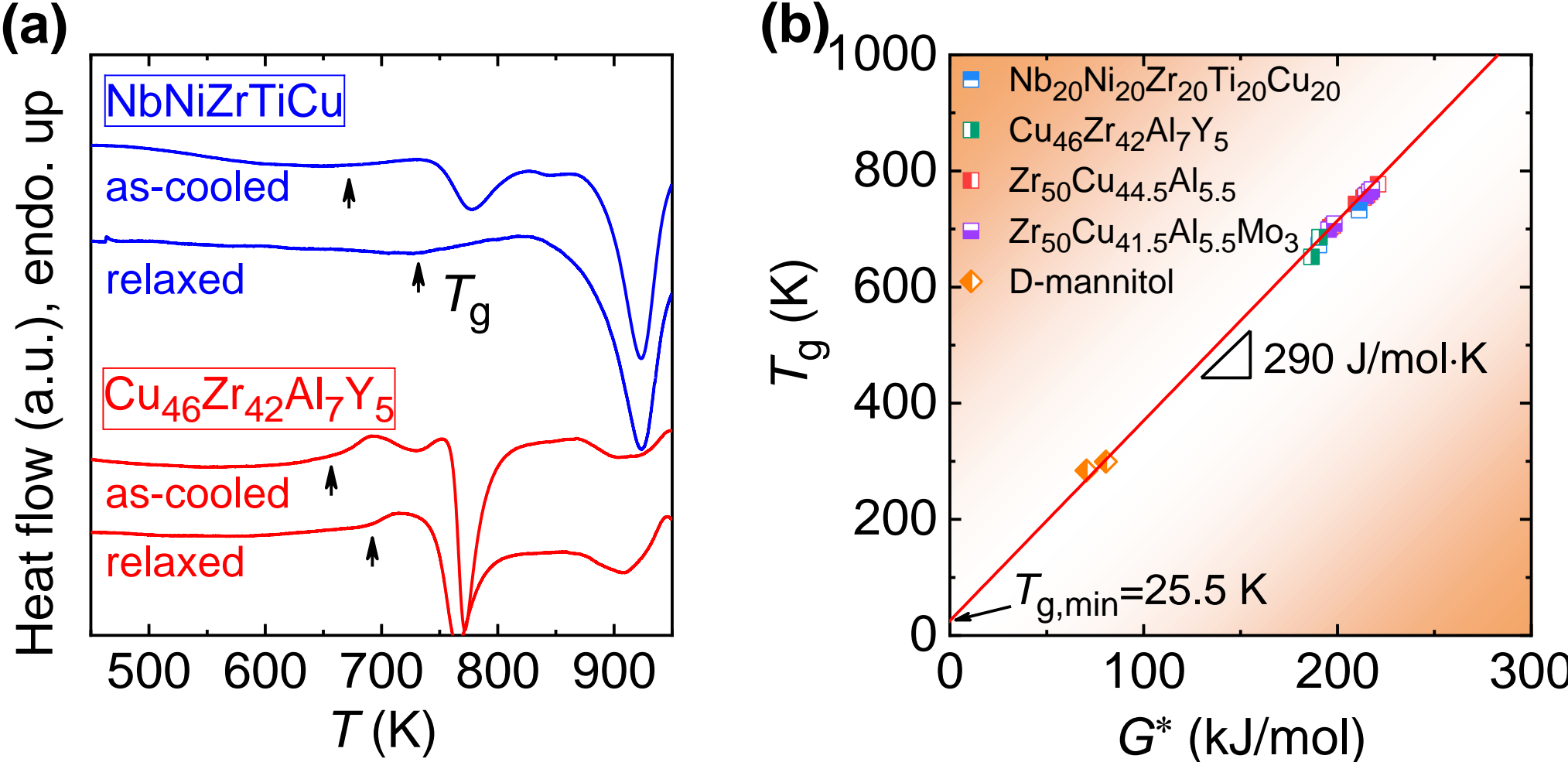

Fig. 2. The calorimetric glass transition in polyamorphic metallic glasses and its relation to the activation free energy $G^*$. (a) The representative DSC curves for both the as-cooled metallic glass and the relaxed high-stable glass with higher $T_g$. The upper two traces are obtained from high-entropy NbNiZrTiCu glass. The lower two traces are from $Cu_{46}Zr_{42}Al_7Y_5$ glass. (b) The activation free energy ($G^*$) versus $T_g$ for the polyamorphic glasses. The data for the relaxed $Zr_{50}Cu_{44.5}Al_{5.5}$ and $Zr_{50}Cu_{41.5}Al_{5.5}Mo_3$ glasses are taken from Ref[43], and the data for D-mannitol glasses are from Refs[65,70].

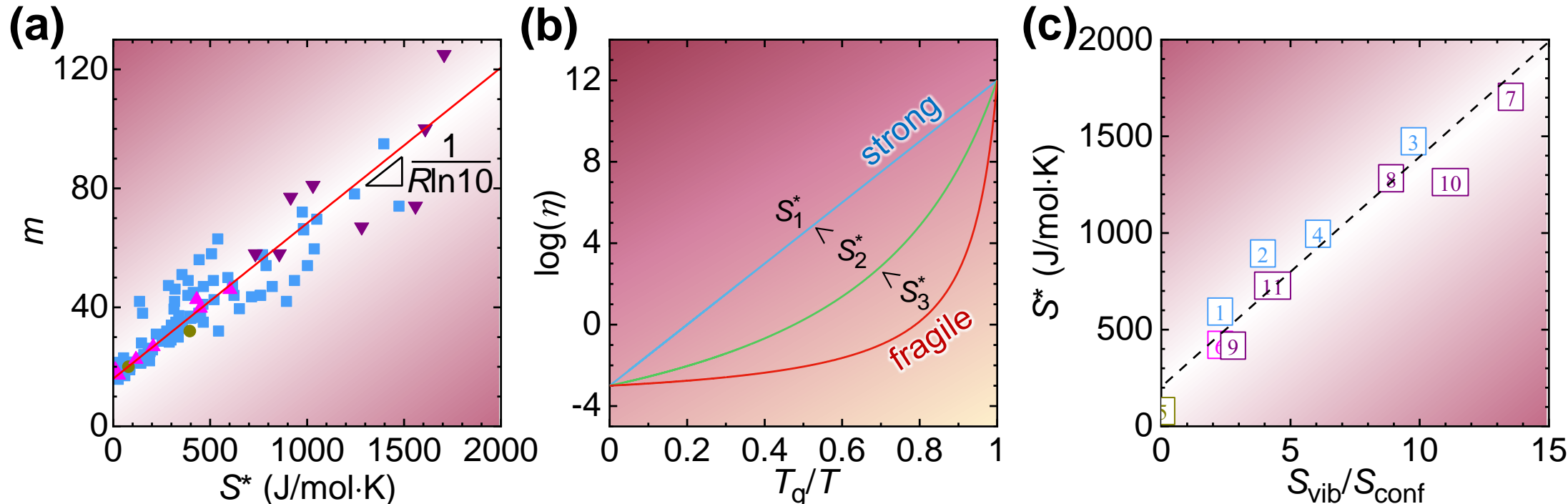

Fig. 3. The correlation between fragility and activation entropy. (a) Fragility $m$ versus the activation entropy $S^*$, which exhibits a linear relationship with a slope of $1/R\ln 10$ ($R$ is gas constant). (b) The schematic Angell plot of viscosity $\eta$ versus $T_g/T$. The strong supercooled liquid has smaller activation entropy than the fragile supercooled liquids. (c) The activation entropy $S^*$ is correlated with the ratio between vibrational entropy $S_{vib}$ and configurational entropy $S_{conf}$. 1:$Pd_{40}Ni_{40}P_{20}$, 2:$Zr_{55}Cu_{30}Al_{10}Ni_5$, 3:$Cu_{50}Zr_{50}$, 4:$Cu_{46}Zr_{46}Al_8$, 5:$SiO_2$, 6:$Ge_{20}Se_{80}$, 7:Toluene, 8:Ethylbenzene, 9:1-Butene, 10:OTP 11:3-Bromopentane.

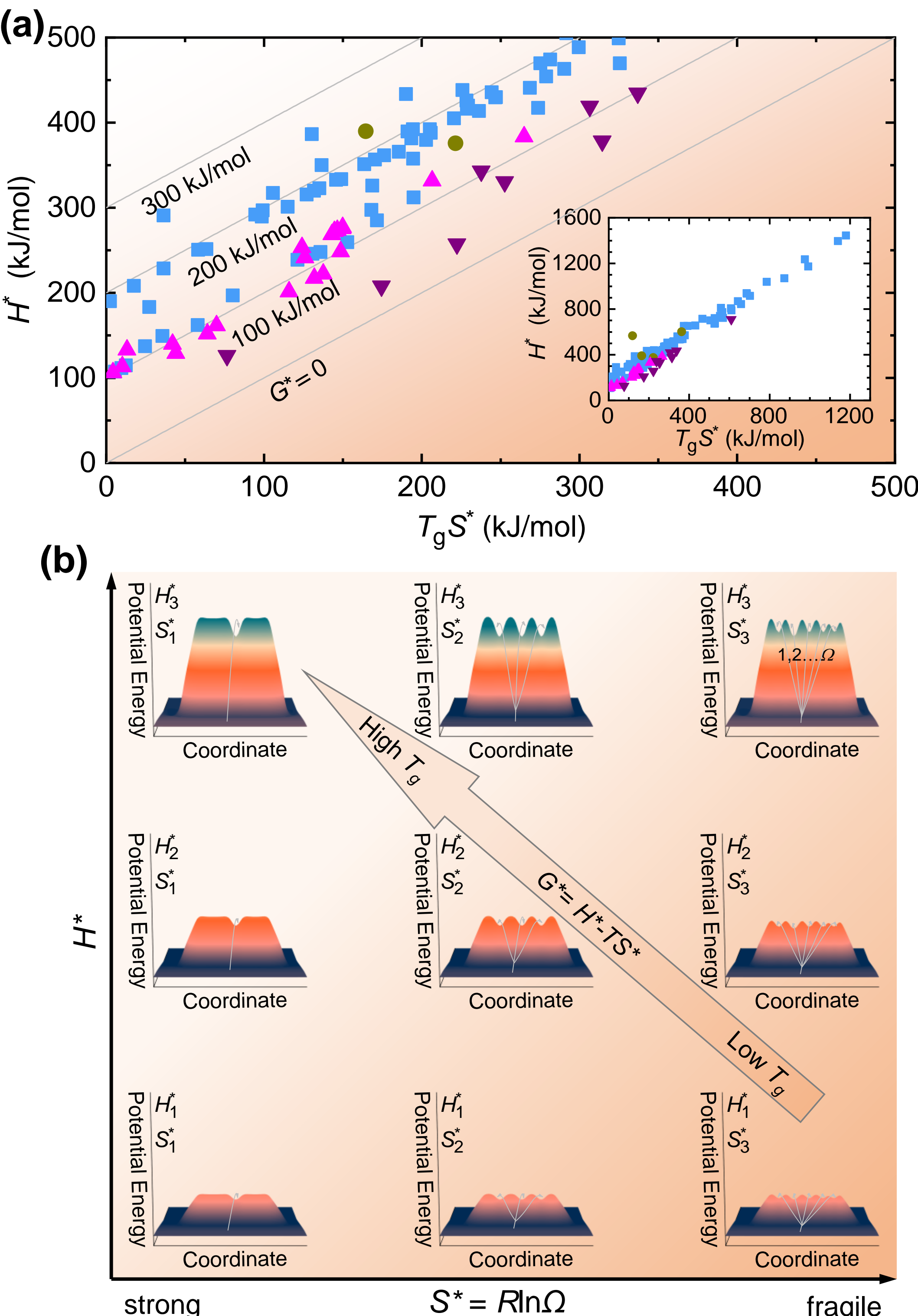

Fig. 4. The kinetic nature of glass transition. (a) The relationship between activation enthalpy and activation entropy. The contour lines of different $G^*=H^*-TS^*$ represent different $T_g$. (b) Schematic potential energy landscape with different activation enthalpy

$H^*$, activation entropy $S^*$ and activation free energy $G^*$. $H^*$ represents the barrier height, $H_1^* < H_2^* < H_3^*$. $S^*$ represents the number/degeneracy ($\Omega$) of transition states with $S^*=R\ln\Omega$, $S_1^* < S_2^* < S_3^*$. $S^*$ determines the fragility of glass forming liquid. $G^*$ determines the $T_g$.